\documentclass[11pt,a4paper]{article} 
\usepackage{jcappub} 
\usepackage{natbib}
\usepackage{hyperref}

\newcommand{\cmt}[1]{}
\newcommand{\Rs}{R_{sub}}

\newcommand{\de}{\partial}
\newcommand{\pj}{p_{inj}}
\newcommand{\xij}{\xi_{inj}}

\newcommand{\Rtt}{R_{tot}}

\def\cm3{~{\rm cm^{-3}}}
\newcommand{\rxj}{RX J1713.7-3946}

\def\gtrsim{\buildrel > \over {_{\sim}}}

\title{
Cosmic-ray acceleration in supernova remnants: non-linear theory revised
}

\author{Damiano Caprioli}

\affiliation{
Department of Astrophysical Sciences, Princeton University, Princeton, NJ 08544, USA
}
\emailAdd{caprioli@astro.princeton.edu}

\abstract{
A rapidly growing amount of evidences, mostly coming from the recent gamma-ray observations of Galactic supernova remnants (SNRs), is seriously challenging our understanding of how particles are accelerated at fast shocks.
The cosmic-ray (CR) spectra required to account for the observed phenomenology are in fact as steep as $E^{-2.2}$--$E^{-2.4}$, i.e., steeper than the test-particle prediction of first-order Fermi acceleration, and significantly steeper than what expected in a more refined non-linear theory of diffusive shock acceleration. 
By accounting for the dynamical back-reaction of the non-thermal particles, such a theory in fact predicts that the more efficient the particle acceleration, the flatter the CR spectrum.
In this work we put forward a self-consistent scenario in which the account for the magnetic field amplification induced by CR streaming produces the conditions for reversing such a trend, allowing --- at the same time --- for rather steep spectra and CR acceleration efficiencies (about 20\%) consistent with the hypothesis that SNRs are the sources of Galactic CRs. 
In particular, we quantitatively work out the details of instantaneous and cumulative CR spectra during the evolution of a typical SNR, also stressing the implications of the observed levels of magnetization on both the expected maximum energy and the predicted CR acceleration efficiency.
The latter naturally turns out to saturate around 10-30\%, almost independently of the fraction of particles injected into the acceleration process as long as this fraction is larger than about $10^{-4}$.
}

\begin{document}
\maketitle

\section{Cosmic-ray spectra: need for steep}
According to the well-known \emph{supernova paradigm}, the bulk of the cosmic rays (CRs) we detect at Earth in a vast range of energies spanning from about a GeV to about $10^8$GeV is of Galactic origin, and supernova remnants (SNRs) are the best candidates as acceleration places.

The reason of the popularity of such a scenario is indeed twofold: on one hand, as already put forward many decades ago \cite{baade-zwicky34}, this class of astrophysical objects can account for the required CR energetics and, on the other hand, a very general and efficient mechanism \cite{Fermi54} turned out be very effective for accelerating particles diffusing around SNR forward shocks \cite{krymskii77, axford+77,bell78a,blandford-ostriker78}.

Such a mechanism, which is usually referred to as \emph{diffusive shock acceleration} (DSA), is particularly appealing because it naturally predicts the energy spectrum of the accelerated particles to be a power-law $\propto E^{-q}$, whose spectral index $q=\frac{r+2}{r-1}$ does not depend on the microphysics of the scattering processes but only on the compression ratio $r$ felt by diffusing particles.
Moreover, since for any strong (i.e., with sonic Mach number much larger than 1) shock $r=4$, it is straightforward to predict SNR strong shocks to accelerate particles with spectral index  $q=2$.
 
For many years, the most prominent evidence of the non-thermal activity ongoing in SNRs has been the detection of synchrotron emission in the radio band due to relativistic electrons \cite[e.g.][]{case-bhattacharia98}, lately corroborated by X-ray observations assessing the extension of the electron spectrum up to multi-TeV energies \citep{koyama+95}.

Finally, in the last few years, the present generation of $\gamma$-ray instruments has opened an additional window on the SNR phenomenology in terms of non-thermal content: the high-quality data from GeV-range satellites (AGILE and Fermi), and from TeV-range ground-based Cherenkov telescopes (CANGAROO, HESS, VERITAS, MAGIC,...) provided us with unprecedented insights into the physical processes involving both accelerated leptons and hadrons.
We have in fact objects, like SNR \rxj, whose $\gamma$-ray emission is likely due to very energetic electrons radiating through inverse-Compton scattering against some photon background \cite{Fermi1713}, and objects like Tycho's SNR, whose emission is likely due to the decay of neutral pions produced in nuclear collisions between relativistic nuclei (with energies as high as $\sim$500 TeV, at least) and the background plasma \citep{Tycho}. 
With a terminology that has become rather popular, we refer to these scenarios as \emph{leptonic} or \emph{hadronic} depending on the population of non-thermal particles mainly responsible for the observed $\gamma$-ray emission.

This brand new, and constantly increasing, phenomenology of SNRs can significantly probe our comprehension of how DSA works in this class of sources, and even further information is expected to come from the next generation of $\gamma$-ray experiments, currently represented by the \emph{Cherenkov Telescope Array} \citep{CTA}.

As pointed out for instance in ref.~\cite{gamma}, the photon spectral index in most of the $\gamma$-ray bright SNRs is inferred to be appreciably larger than 2.
While in the TeV range this may be due to the fact that we are dealing with the cut-off of the parent particle distribution (in both scenarios), a rather steep spectrum in the GeV range cannot be ascribed to inverse-Compton scattering (which would return a photon spectrum $\propto E^{-1.5}$ for a $E^{-2}$ electron one), therefore representing a strong signature of hadron acceleration.

In this paper we do not want to address the question whether most of the SNR $\gamma$-ray emission is either of hadronic or of leptonic origin, but we want instead to stress a point which, in our opinion, has been given too little attention in the literature: $\gamma$-ray observations are building up more and more evidences that SNRs accelerate particles with spectra different, and namely steeper, than what predicted by test-particle DSA at strong shocks.

In the hadronic scenario, in fact, the $\gamma$-ray spectrum has to be parallel to the one of parent hadrons.
In the leptonic scenario, instead, in order to be consistent with the steep spectra ($\propto E^{-2.2}$--$E^{-2.5}$) observed in the GeV band, a contribution from non-thermal bremsstrahlung from $\sim$GeV electrons must be added to the inverse-Compton one to fit the data \citep[see, e.g., the case of Cas A in ref.][]{CasAFermi}. 
Also for bremsstrahlung emission, however, the photon spectrum is parallel to the one of the parent particles, in turn implying that also GeV electrons must have a spectrum steeper than $E^{-2}$.
It is worth recalling that in this region of the electron spectrum no cooling via synchrotron emission is effective, therefore protons and electrons are expected to show the same spectral index.

Quite intriguingly, for about 30 years scientists have been working out the extension of DSA to the case in which CRs are not simply test-particles but instead participate actively in the shock dynamics, carrying sizable fractions of pressure and energy density of the system \citep[see, e.g., refs.][for thorough reviews on the topic]{drury83, blandford-eichler87, jones-ellison91, malkov-drury01}.
Such a \emph{non-linear theories of DSA} (usually referred to as NLDSA) invariably predicts the back-reaction of the accelerated particles to induce in the upstream the formation of a precursor in which the fluid is slowed down because of the pressure in CRs diffusing around the shock.
The net result is that particles with larger momenta, and in turn larger diffusion lengths, ``feel'' --- on average --- fluid compressions larger than 4.
On the contrary, particles with mildly suprathermal momenta only ``see'' a weaker subshock, with a compression ratio smaller than 4.
This spread in the fluid compression ratio experienced by CRs leads to rather concave spectra, steeper (flatter) than $E^{-2}$ at low (high) energies. 
Moreover, the standard NLDSA theory predicts such a concavity to become more and more marked when the acceleration efficiency increases, so that fractions of 10--50\%  of the shock ram pressure channeled into CRs naturally imply spectra as flat as $E^{-1.7}$--$E^{-1.5}$ \emph{above a few GeV} \citep{comparison}.
The steepening induced by the fluid precursor can in fact be effective only \emph{below} this threshold, GeV-nuclei being the carriers of most of the CR pressure when the spectrum is steeper than $E^{-2}$.
Since the photons produced in nuclear collisions have a typical energy about 10 times lower than the parent hadron, sub-GeV to multi-TeV observations invariably probe energy regions where standard NLDSA theory predict spectra systematically flatter than $E^{-2}$.

No clear-cut evidence of concavity has been found in any of the $\gamma$-ray-bright SNRs, but this may easily be a consequence of the intrinsic errors in the measurements, which cannot be more accurate than a 10--20\% in most cases.
What was really unexpected, though, is that all of the observed spectra \citep[see data collected in ref.][]{gamma} are either consistent with, or steeper than, the test-particle prediction $\propto E^{-2}$ and not with NLDSA ones!

CR spectra as steep as $E^{-2.3}$--$E^{-2.4}$ are also required when trying to link two distinct but actually related aspects of the SNR paradigm, namely the spectra before (at injection) and after (at Earth) propagation in the Milky Way. 
The widely accepted model for CR transport in the Galaxy, mainly based on CR isotope and secondary-to-primary compositions, suggests the residence time in the Galaxy to scale as $E^{-\delta}$, with $\delta\sim$0.3--0.6 \citep{galprop,dragon08,ba11a}.
The CR flux observed at Earth ($\propto E^{-2.75}$) has in fact to be proportional to the injection spectrum $\propto E^{-q}$ multiplied by the Galactic residence time $\propto E^{-\delta}$, providing the constraint $q+\delta\simeq 2.75$ and therefore implying $q=$2.2--2.4. 
In addition, the smallest value $\delta\simeq 0.3$ is preferred in order to account for the relatively small anisotropy observed in the direction of arrival of CRs above $\sim 1$ TeV \citep[see, e.g.,][]{ba11b}.
Again, a hint that the spectra of CRs accelerated in SNRs have to be non-negligibly steeper than the test-particle case, and significantly steeper than standard NLDSA predictions. 

In this respect, it is however important to remember that the spectrum injected into the Galaxy during the whole SNR lifetime has not to be compared one-to-one with $\gamma$-ray observations, since such a spectrum is expected to be a rather complex superposition of several time-dependent contributions, as illustrated in ref.~\citep{crspectrum}. 
Despite of our still incomplete understanding of how accelerated particles leave their source and become CRs, physically motivated calculations \cite{bv07,pzs10,nuclei} indicate that the cumulative spectrum (i.e., integrated over the SNR lifetime) turns out to be only $\sim 0.05$--0.1 steeper in slope than the instantaneous spectrum achieved in the early Sedov stage.

We can safely summarize the points above by saying that the observational evidences are better recovered if the spectra of accelerated particles at SNR shocks were steeper than the test-particle prediction $\propto E^{-2}$.
How this is actually achieved is a fundamental and challenging question whose answer cannot be found in standard NLDSA theories, since they invariably predict that the more efficient the acceleration, the flatter the spectrum of accelerated particles. 

Therefore, we are left with an open dilemma: \emph{are Galactic CRs efficiently accelerated in SNRs?} 
And in this case, \emph{what is the physical mechanism able to reverse the standard prediction of the NLDSA theory?}
Understanding why observed $\gamma$-ray spectra from SNRs are so steep is indeed a fundamental tile in the comprehension of particle acceleration at shocks, and not an almost negligible correction to an otherwise solid physical theory.

In this paper we investigate one possible mechanism, put forward already at the dawn of DSA theories \citep{bell78a}, which, in a revised form accounting also for the more recent evidences of magnetic field amplification in young SNRs, has been proposed as a viable scenario for accelerating particles with rather steep spectra consistent with $\gamma$-ray observations \citep{gamma}.
The main idea here is that, since efficient CR acceleration may induce a very efficient amplification of the magnetic field via some plasma instabilities, the magnetic structures acting as scattering centers for the CR diffusion may achieve a non-negligible velocity with respect to the background fluid \citep[also see refs.][for other recent implementation of the same effect]{zp08b,jumpkin,kang-ryu10,lee-ellison12}. 
This phenomenon may significantly alter the actual compression ratio felt by accelerated particles and, in turn, their spectrum and eventually the global shock dynamics.

The original contribution of this work is a quantitative analysis of such an effect, assessing --- as self-consistently as possible --- the spectral steepening as a function of the CR acceleration efficiency, here regulated by the number of particle extracted by the thermal bath and injected in the acceleration process.

The plan of the paper is as follows: in section \ref{sec:model} the details of the model are illustrated.
The solutions of the equation describing the CR transport and the saturation of the magnetic field amplification are worked out, along with their non-linear interplay with the SNR hydrodynamics.
In section \ref{sec:results} we discuss our main results, namely the fact that including the finite velocity of the scattering centers in the amplified magnetic field it is possible to reproduce many observational features, like the levels of magnetic field amplification, the steepness of the CR spectra and the achievement of maximum energies consistent with the knee.
We also outline the relationships between instantaneous and cumulative spectra in different stages of the SNR evolution and, quite interestingly, point out for the first time how our findings does not depend in a critical way on the fraction of particle injected, as long as this is larger than $\sim 10^{-4}$.
After some comments about strong and weak points of the present model, we conclude in section \ref{sec:conclusions}.

\section{The kinetic model for particle acceleration}\label{sec:model}
The final goal of any approach to the problem of NLDSA is to solve self-consistently the equations for mass, momentum and energy conservation, along with a description of the non-thermal particles and, possibly, their generation of magnetic waves via plasma instabilities.

Apart from computationally very expensive particle-in-cell (PIC) simulations of collisionless shocks, in which the interplay between particles and fields is calculated from first principles, the most common way to describe the transport of relativistic particles is either to prescribe a scattering law \citep[Monte Carlo approaches, see for instance refs.~][]{jones-ellison91,ebj95,veb06} or to solve a Vlasov-like diffusion-convection equation for the isotropic part of the CR distribution function \cite{skilling75a}. 
More precisely, the diffusion-convection equation has been solved numerically in its time-dependent formulation \citep{bell87, bv97, kang-jones97, KJG02, kang-jones05, za10} or semi-analytically, i.e., by integrating the quasi-stationary equation in order to have an implicit analytical expression for the CR distribution function \citep{malkov97,mdv00, blasi02, blasi04, ab05, boundary}.
Nevertheless, all of the different approaches involving a model for the particle transport lead to very consistent results for non-relativistic shocks, as demonstrated in ref.~\cite{comparison}, and at the moment represent our best description of particle acceleration in SNRs.

In this work we apply the semi-analytical formalism for NLDSA put forward in refs.~\cite{ab05,ab06}, in its extended version including particle escape from an upstream boundary \citep[][]{boundary} and the dynamical feedback of self-generated magnetic fields \citep[][]{jumpl,jumpkin}.
Again, we would like to stress that the limitations and the uncertainties in applying this NLDSA model to SNR shocks are shared by virtually \emph{any} non-PIC approach: the main advantage of a semi-analytical formalism is to be very quick and versatile (a run takes several seconds on a standard laptop), and therefore well-suited for investigating a large parameter space.

The stationary mass and momentum conservation equation for a plane, non-relativistic shock simply read
\begin{eqnarray}
\rho(x)u(x)&=&\rho_0 u_0\label{eq:mass}\\
\rho(x)u(x)^2+ p(x)+ p_{cr}(x)+p_B(x)&=& \rho_0u_0^2+p_{g,0}+p_{B,0},\label{eq:momentum}
\end{eqnarray}
where $\rho,u$ and $p$ represent the plasma density, velocity (in the shock reference frame) and pressure, while the subscripts $cr$ and $B$ label the pressure in the shape of non-thermal particles and magnetic fields.
Throughout the paper the subscripts 0, 1 and 2 refer to physical quantities measured at upstream infinity, immediately upstream of the shock and downstream, respectively. The shock is stationary at $x=0$ and, by definition, $u_0=V_{sh}$, i.e., the shock velocity in the Earth reference frame.

Global conservation of energy is then granted by taking into account the proper jump conditions at the shock \citep{jumpl} and by solving the space-dependent equations for the energy transport of thermal gas, magnetic waves and CRs.
In particular, we consider the upstream thermal plasma to be adiabatic, namely
\begin{equation}\label{eq:gas}
\frac{p(x)}{\rho(x)^{\gamma}} = \frac{p_0}{\rho_0^{\gamma}}; \quad \gamma=\frac{5}{3},
\end{equation}
\citep[a generalization to non-adiabatic heating is straightforward, see ref.][]{jumpkin}
and we model the growth and the advection of the Alfv\'en modes generated by accelerated particles via streaming instability by solving the transport equation for the magnetic turbulence \citep[e.g.,][]{mckenzie-volk82} according to the fluid formalism in ref.~\cite{jumpkin}:
\begin{equation}\label{eq:wave}
2 \tilde{u}(x)\frac{d p_B(x)}{d x}= v_A(x)\frac{d p_{cr}(x)}{d x}-3 p_B(x) \frac{d \tilde{u}(x)}{d x}.
\end{equation}
This stationary equation simply describes the advection of the magnetic pressure (left-hand side) as produced by the CR pressure gradient (first term on the right-hand side), also accounting for adiabatic compression in the precursor (second term on the same side).
Here we consider Alfv\'en modes generated by CR streaming instability and thus propagating against the fluid, therefore $\tilde{u}(x)=u(x)+v_A(x)$ is the fluid velocity in the wave reference frame, where particle scattering is elastic. 

We will comment the generality of our choice more widely in the following sections, but it is worth stressing that eq.~\ref{eq:wave}, in its simplicity, still captures a basic feature of the interplay between CRs and magnetic field: the CR diffusion velocity $\sim\frac{D}{p_{cr}}\frac{dp_{cr}}{dx}\sim \tilde{u}$ is typically larger than the Alfv\'en velocity, therefore the wave-particle coupling tends to restore an equilibrium between the two by making $v_A$ larger and the gradient in $p_{cr}$ smaller, i.e., by amplifying the magnetic field and smoothing the precursor.

Finally, we include the diffusion-convection equation for the isotropic part of the CR distribution function $f(x,p)$ \citep[see, e.g., ref.][]{skilling}:
\begin{equation}
\tilde{u}(x)\frac{\de f(x,p)}{\de x}=\frac{\de}{\de x}\left[D(x,p)\frac{\de f(x,p)}{\de x}\right]+\frac{p}{3}\frac{{\rm d} \tilde{u}(x)}{{\rm d} x}\frac{\de f(x,p)}{\de p}+Q(x,p),
\label{eq:trans}
\end{equation}
where $Q(x,p)$ accounts for particle injection and 
\begin{equation}
D(x,p)=\frac{v(p)}{3}r_L(x,p)
\end{equation}
is the Bohm-like parallel diffusion coefficient for a particle with velocity $v(p)$ and Larmor radius $r_L(x,p)=\frac{pc}{eB(x)}$ in the local, amplified magnetic field $B(x)$.

Following the implementation in ref.~\cite{bgv05}, we assume that, immediately behind the shock, all the particles in the Maxwellian tail with a momentum larger than a critical $\pj$ have a Larmor radius large enough to be able to cross the shock and return upstream \citep[\emph{thermal leakage}, see e.g.,][]{KJG02}.
Such an injection momentum is parametrized as a multiple of the downstream thermal momentum $p_{th,2}$, namely 
\begin{equation}
\pj=\xij p_{th,2};\qquad p_{th,2}=\sqrt{2m_p k_B T_2},
\end{equation}
where $m_p$ is the proton mass, $k_B$ the Boltzmann constant and $T_2$ the downstream temperature.
We thus have  
\begin{equation}\label{eq:Q}
Q(x,p)=\eta\frac{\rho_{1}u_{1}}{4\pi m_p \pj^2}\delta(p-\pj)\delta(x)\,,
\end{equation}
where 
\begin{equation}\label{eta}
	\eta=\frac{4}{3\sqrt{\pi}}(\Rs-1)\xij^3 e^{-\xij^2}\,,
\end{equation}
represents the fraction of the particles crossing the shock injected in the acceleration process and $\delta(x)$ accounts for injection to occur at the shock position. 
Since the additional pressures in CRs and magnetic fields induce a velocity gradient in the upstream (precursor), it is convenient to introduce two distinct compression ratios \emph{as felt by scattered particles}, the subshock and the total ones:
\begin{equation}\label{eq:rsrtt}
R_{sub}=\frac{u_{1}+v_{A,1}}{u_{2}};\qquad R_{tot}=\frac{u_{0}+v_{A,0}}{u_{2}}.
\end{equation}
These compression ratios differ from the fluid ones because of the finite velocity of the scattering centers $v_A$, which vanishes downstream because of wave isotropisation ($v_{A,2}\approx0$).
For typical interstellar values also $v_{A,0}/u_0\ll 1$, but when efficient magnetic field amplification occurs in the precursor $v_A$ may become a non-negligible fraction of $u_1$ \citep[also see the discussion in ref.][]{jumpkin}.
Moreover, when self-generated by CRs, waves tend to propagate in such a way to smooth the CR pressure gradient out, i.e., \emph{against} the fluid: very generally, we have $\tilde{u}=u+v_A\leq u$, so that the compression ratios in eq.~\ref{eq:rsrtt} turn out to be \emph{smaller} than their fluid counterparts (see section \ref{sec:comments} for additional comments).

The solution of eq.~(\ref{eq:trans}) with the spatial boundary condition $f(x_{0},p)=0$, which mimics the presence of an upstream free escape boundary placed at $x=x_{0}$, can be written as \citep{boundary}: 
\begin{eqnarray}\label{eq:app}
f(x,p)&=&f_{2}(p)\exp\left[-\int_{x}^{0} dx'\frac{\tilde{u}(x')}{D(x',p)}\right]	\left[ 1-\frac{W(x,p)}{W_0(p)}\right];\\
\Phi_{esc}(p)&=&-D(x_0,p)\left. \frac{\partial f}{\partial x}\right|_{x_0}=- \frac{u_0f_{2}(p)}{W_{0}(p)}\,;\\
	W(x,p)&=&\int_{x}^{0} dx'\frac{u_0} {D(x',p)}\exp\left[\int_{x'}^{0}dx''\frac{\tilde{u}(x'')}{D(x'',p)}\right].
\end{eqnarray}

Here $\Phi_{esc}(p)$ represents the flux of particles escaping the system at $x_0$ \citep{escape} and the distribution function at the shock $f_2(p)=f_1(p)$ reads:
\begin{equation}\label{eq:solshock}
f_2(p)=\frac{\eta n_{0} q_p(p)}{4\pi \pj^{3}}
	\exp\left\{-\int_{\pj}^{p}\frac{dp'}{p'}
	q_p(p')\left[ U_p(p')+\frac{1}{W_0(p')}\right]\right\},
\end{equation}
where 
\begin{equation}\label{eq:Up}
U_{p}(p)=\frac{\tilde{u}_1}{u_0}-
	\int_{x_{0}}^{0} \frac{dx}{u_0 } 
	\left\{\frac{\partial \tilde{u}(x)}{\partial x}
	\exp\left[-\int_{x}^{0} dx'\frac{\tilde{u}(x')}{D(x',p)}\right]	
		\left[ 1-\frac{W(x,p)}{W_0(p)}\right]\right\}
\end{equation}
and 
\begin{equation}
q_p(p)=\frac{3\Rtt}{\Rtt U_{p}(p)-1}
\end{equation}
correspond to the \emph{average} fluid velocity and spectral slope relevant for a particle with momentum $p$.
In the test-particle limit, in fact, $\frac{\partial u}{\partial x}=0$ and in turn $U_p=1$ and $q_p=\frac{3r}{r-1}$.

Even if the calculations are led in the more natural framework of the \emph{momentum} space for the CR distribution function, $f(p)$, in the rest of the paper we will use the \emph{energy} distribution function $\phi(E)=4\pi p^2 f(p)\frac{dp}{dE}$ to provide a clearer comparison with observations. 
  
\subsection{Magnetic field amplification}\label{sec:MFA}
When particle acceleration is efficient, magnetic field amplification due to CR-induced instabilities may eventually lead to $\delta B/B\gg 1$ non only in the downstream but even in the upstream: such a scenario is motivated, for instance, by the lack of detection of X-ray emission from the precursor of SN 1006 \citep{gio-sn1006}. 
This evidence is to some extent complementary to the well-know evidences of large magnetic fields in the downstream, and strongly suggests that magnetic field amplification is induced by CRs rather than by some process occurring at or behind the shock only. 

Even if the resonant Alfv\'en modes produced by streaming instability \citep{bell78a} are purely transverse, the presence of other non-resonant modes and the likely onset of a certain degree of turbulence might lead to an --- at least partial --- isotropisation of the amplified magnetic field, in turn affecting the component of the field parallel to the shock normal and eventually increasing the effective Alfv\'en velocity, as demonstrated by the authors in ref.~\cite{rs10}.
In order to heuristically take this effect into account, we assume that in eq.~(\ref{eq:wave}) the relevant Alfv\'en velocity is the one in the local amplified field, namely
\begin{equation}
v_A(x)=\frac{B(x)}{\sqrt{4\pi \rho(x)}};\quad B(x)=\sqrt{8\pi p_B(x)}.
\end{equation}
Normalizing the velocities to the shock velocity $u_0$ and the pressures to the ram pressure $\rho_0u_0^2$ (indicated with capital letters) we can write eq.~(\ref{eq:wave}) as 
\begin{equation}\label{eq:wave_norm}
2U(x)\frac{d P_B(x)}{dx}=V_A(x)\frac{dP_{cr}(x)}{dx}-3P_B(x)\frac{dU(x)}{dx}.
\end{equation}
Since $P_B\propto M_A^{-2}$ eqs.~\ref{eq:wave} and \ref{eq:wave_norm} are accurate at the second order in $V_A/U$, therefore the correction due to the finite velocity of the scattering centers in the advective and compressional terms can be neglected and we can approximate, in this equation, $\tilde{U}(x)$ as $U(x)$.

Moreover, since SNR shock typically show both sonic and Alfv\'enic Mach numbers much larger than one, at the order $M_A^2\gg 1$, $M_s^2\gg 1$, from eq.~(\ref{eq:momentum}) we have $P_{cr}\simeq 1- U(x)$ and therefore eq.~(\ref{eq:wave_norm}) becomes:
\begin{equation}
\frac{d P_B(x)}{dx}=-\left[\sqrt{\frac{P_B(x)}{2U(x)}}+\frac{3}{2}\frac{P_B(x)}{U(x)}\right]\frac{dU(x)}{dx}.
\end{equation}
Introducing $\Psi(x)=\sqrt{\frac{P_B(x)}{2U(x)}}$ we get the simple equation 
\begin{equation}
\frac{d\Psi(x)}{dx}=-\frac{5\Psi(x)+1}{4U(x)}\frac{dU(x)}{dx},
\end{equation}
whose solution with the boundary condition $\Psi(x_0)=\sqrt{P_B(x_0)/2}\simeq 0$ is 
\begin{equation}
\Psi(x)=\frac{1}{5}\left[1-U(x)^{-5/4}\right]
\end{equation}
thus finally leading to:
\begin{equation}\label{eq:PB}
P_B(x)=\frac{2}{25}\frac{[1-U(x)^{5/4}]^2}{U(x)^{3/2}}.
\end{equation}

This equation, besides taking into account the adiabatic compression of the modes in the precursor due to the factor $U(x)^{-3/2}$, connects the local amount of magnetic field with the local modification induced by CRs, and therefore naturally predicts the maximum amplification to be achieved immediately upstream of the subshock.
If this were rigorously the case, it would be unlikely to have non-linear magnetic structures with large phase velocities far in the precursor, where the high-energy particles diffuse, and therefore the standard NLDSA prediction for the total compression ratio felt by the highest energy particles would be recovered.

On the other hand, there are many effects that are not included in eq.~(\ref{eq:wave_norm}) which may play a relevant role, like for instance the excitation of Bell's non-resonant modes \citep{Bell04,Bell05}, or the development of large scales structures due to \emph{fire-hose instability} \citep{gary+84,shapiro+98}, which may be most effective where the CR distribution function is most anisotropic, i.e., far in the precursor.
There are also phenomenological reasons for the plausibility of such a scenario: in Tycho's SNRs there are evidences for macroscopical magnetic structures on the scales of the Larmor radius of $\sim10^6$GeV particles \citep{eriksen+11}.

In the absence of a consistent kinetic theory for the CR-magnetic field interplay in the precursor, we will assume in the rest of the paper a constant magnetic field in the upstream, whose strength is given by the saturation at the shock from eq.~(\ref{eq:PB}), a choice that proved itself to be adequate in particular for the account of Tycho's multi-wavelength emission and hydrodynamics \citep{Tycho}.
We will further comment on this important point in section \ref{sec:comments}.

Eq.~(\ref{eq:PB}), coupled with eq.~(\ref{eq:gas}) for the background plasma and the solution of eq.~(\ref{eq:trans}) for the CRs, allows for global energy conservation.
The connection between subshock and total compression ratios for the fluid can be finally worked out from the solution of the Rankine--Hugoniot at the subshock also including the dynamical effect of the amplified magnetic field according to equation 16 in ref.~\cite{jumpkin}.

A full solution of the problem of NLDSA can be obtained by solving iteratively, at any given time,the system of equations including also eqs.~(\ref{eq:mass}) and (\ref{eq:momentum}) through the procedure put forward  in ref.~\cite{boundary}.

\subsection{SNR evolution}
\begin{figure}
\begin{center}
\includegraphics[width=0.9\textwidth]{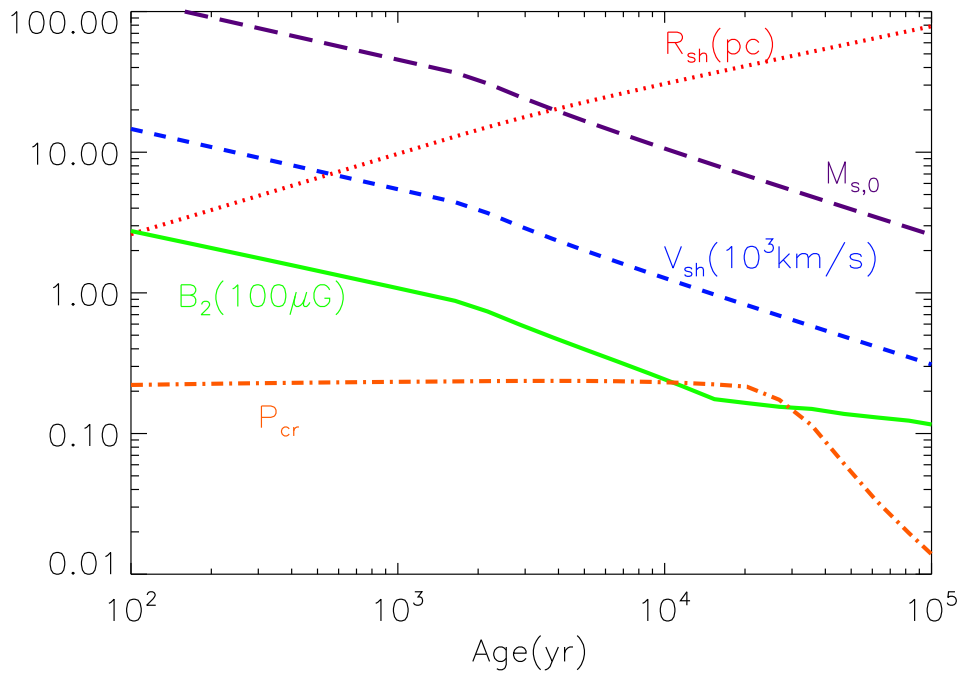}
\caption{Time evolution of several physical quantities: shock radius, velocity and sonic Mach number ($R_{sh}$,$V_{sh}$ and $M_{s,0}$), downstream magnetic field ($B_2$) and CR acceleration efficiency at the shock ($P_{cr}$) for $\xij=3$, $n_0=0.01{\rm cm}^{-3}$ and $T_0=10^6$K. With these environmental parameters the Sedov--Taylor stage begins around 2000 yr and the sonic Mach number drops below 10 around $10^4$yr.}
\label{fig:Hydro}
\end{center}
\end{figure}	

The SNR evolution is modeled following the analytical recipe given in ref.~\cite{TMK99}, and in this paper we consider the propagation of the forward shock in a homogeneous medium with particle density $n_0=\rho_0/m_p=0.01{\rm cm}^{-3}$ and temperature $T_0=10^6{\rm K}$ except when otherwise stated. 
The total kinetic energy and mass in the ejecta is fixed in $E_{SN}=10^{51}$erg and $M_{ej}=1.4M_\odot$, with $M_\odot$ the solar mass. 
The generalization to a more complex circumstellar environment including winds and bubbles produced in the pre-SN stages is rather straightforward \citep{gamma}, but it is omitted here for clarity's sake.

With the environmental parameters chosen the forward shock dynamics is described by:
\begin{equation}
R_{sh}(\tau)\simeq 14.1\,\tau^{4/7}\,{\rm pc};\quad
V_{sh}(\tau)\simeq 4140\,\tau^{-3/7}\,\frac{{\rm km}}{{\rm s}}
\end{equation}
during the ejecta-dominated stage ($\tau=t/T_{ST}\leq 1$), with $T_{ST}\simeq1900$yr, and by 
\begin{equation}
R_{sh}(\tau)\simeq 16.2\left(\tau-0.3\right)^{2/5}\,{\rm pc};\quad
V_{sh}(t)\simeq 3330\left(\tau-0.3\right)^{-3/5}\,\frac{{\rm km}}{{\rm s}}
\end{equation}
during the Sedov--Taylor stage ($\tau=t/T_{ST}\geq 1$), as in figure \ref{fig:Hydro}.
The quasi-stationary solution is calculated at many times from $\tau=0.03$ to the end of the Sedov stage, namely $\tau\sim 60$ according to eq.~13 in ref.~\cite{TMK99}.
The spectra of advected and escaped particles are obtained by convoluting the instantaneous spectra weighted with the shell volumes competing to each time step as described in ref.~\cite{crspectrum}.
Also adiabatic losses due to the shell expansion are taken into account: more precisely, at any time $t\geq t_0$ the energy $E(t)$ of a particle with energy $E_0$ advected downstream at time $t_0$ reads \cite[section  3 of ][]{crspectrum}
\begin{equation}\label{eq:adlosses}
E(t)=E_{0}\left[\frac{V_{sh}(t)}{V_{sh}(t_0)}\right]^{\frac{2}{3\gamma}},
\end{equation}
with $4/3\leq\gamma\leq 5/3$ and a rather week dependence on $\gamma$.

Within this framework, we can study the SNR evolution in terms of particle acceleration across the transition between the ejecta-dominated and the Sedov--Taylor stages, when the CR maximum energy is achieved \citep{bac07}.

\section{A modern view of CR-modified shocks}\label{sec:results}

In this section we want to quantitatively investigate the non-linear response of the system to the presence of accelerated particles and magnetic fields. 
In order to do this, we also need to specify the fraction of particles injected in the acceleration process, $\eta$.
This quantity has not been worked out from first principles, yet, and the only insights come from PIC simulations of collisionless shocks in given regimes \citep[e.g.,][]{ss11,gs12}.
Since in our framework, as in any other approach aimed to account for the temporal and spatial scales relevant to SNRs, we do not have any constraint on the microphysics that regulates particle injection, we allow for the variation of $\xij$, actually the only free parameter in our model.

In the context of the thermal leakage model $\eta$ increases for smaller $\xij$, i.e., when the minimum momentum required for a particle to cross the shock from downstream is closer to the thermal momentum (eq.~\ref{eta}).  
More precisely, in this paper we will span the range $\eta\approx 5\times 10^{-7}-5\times 10^{-2}$ by changing  $\xij$ between 4.5 and 2.5.  

\subsection{Instantaneous and cumulative spectra}
\begin{figure}
\begin{center}
\includegraphics[width=0.87\textwidth]{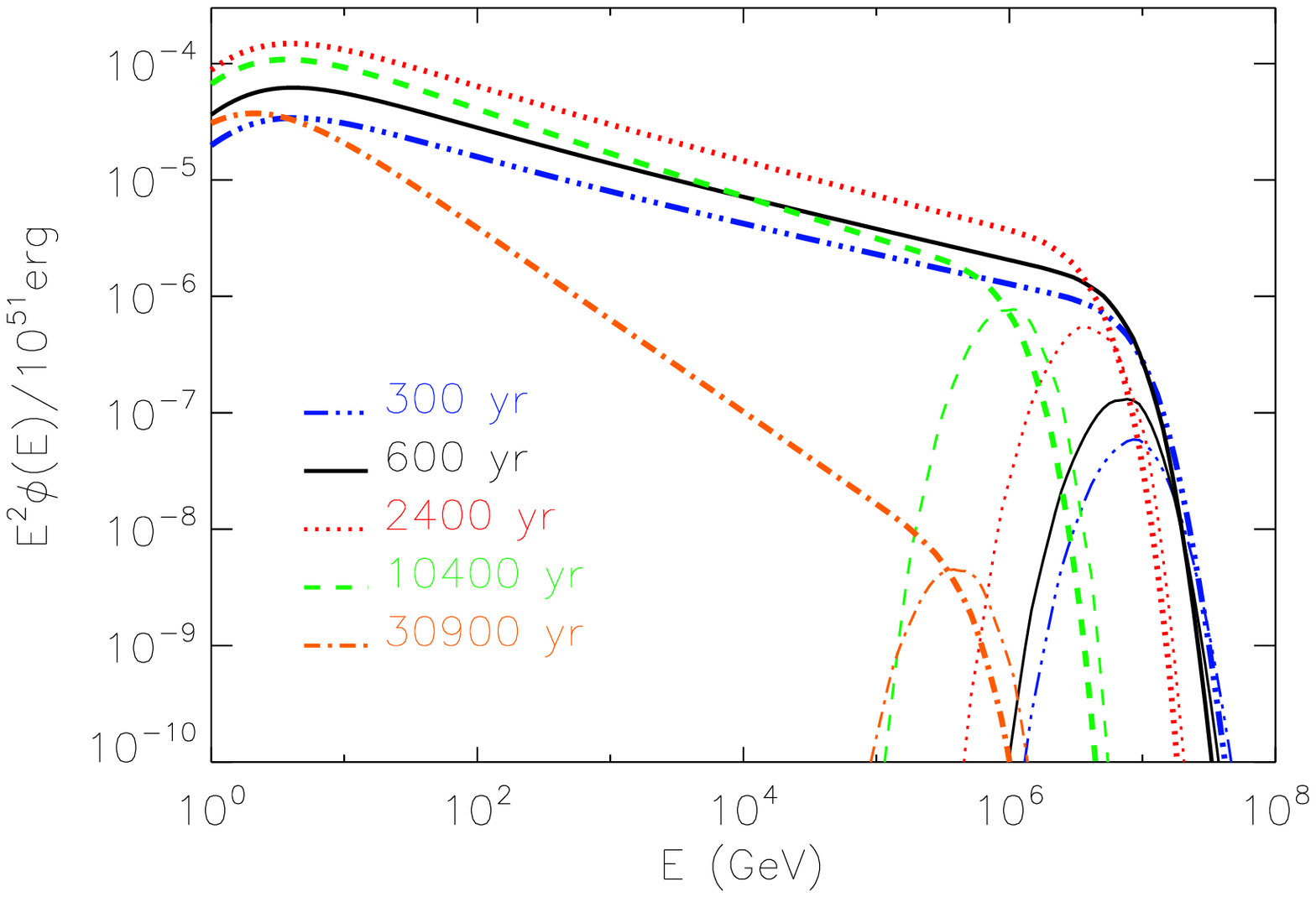}
\includegraphics[width=0.87\textwidth]{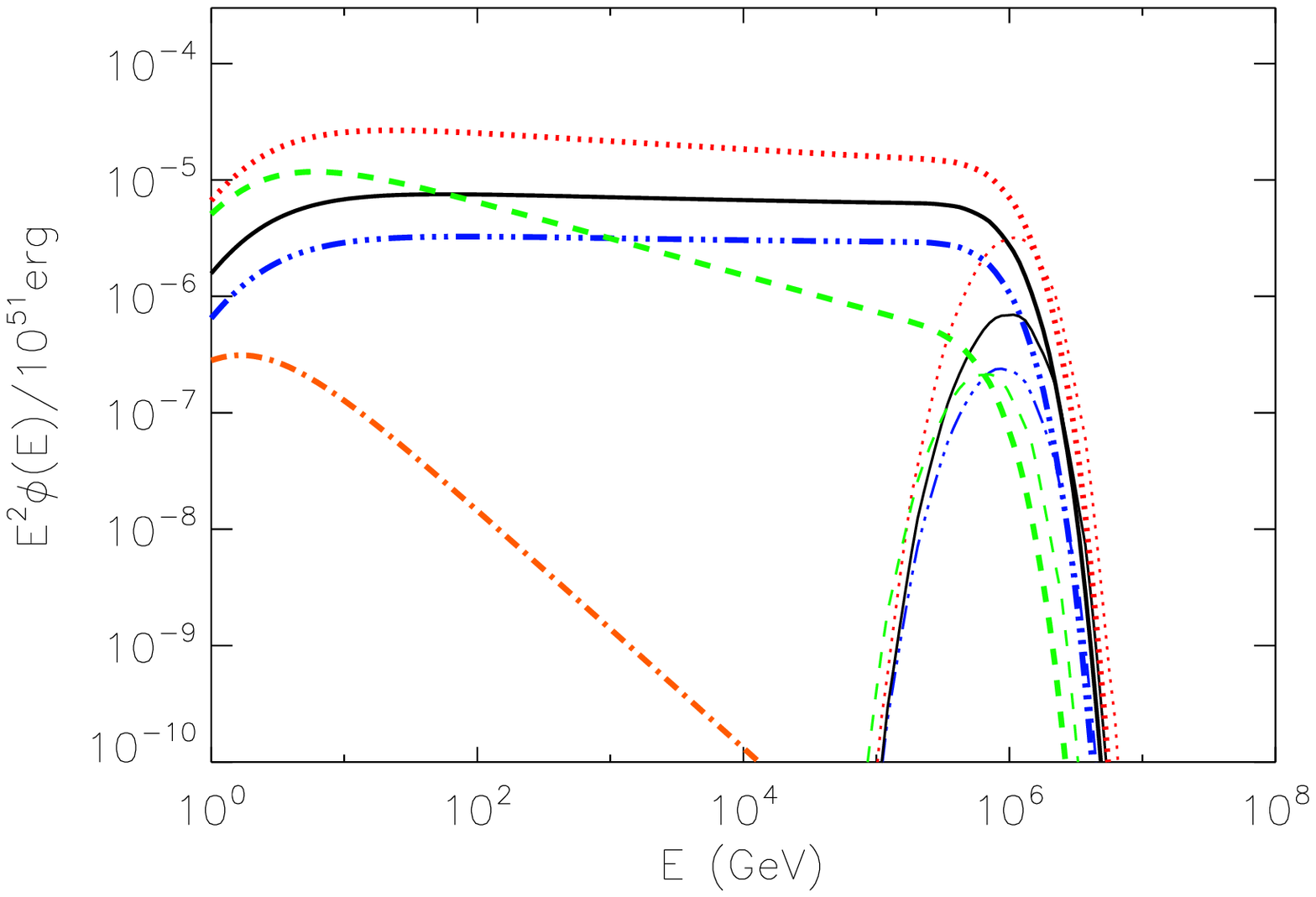}
\caption{Instantaneous CR spectra for different SNR ages as in the legend. 
Thick (thin) lines show the spectra of particles advocated downstream (escaping from the upstream free escape boundary).
The panels correspond to two different injection parameters: $\xij=3$ (top) and $\xij=4$ (bottom panel), corresponding to $\eta\simeq 2.5\times10^{-3}$ and $\eta=4.1\times10^{-6}$, respectively.}
\label{fig:multi}
\end{center}
\end{figure}

The evolution of CR spectra is illustrated in figure \ref{fig:multi}, where the instantaneous spectra of both advected (thick) and escaping particles (thin lines) are shown.  
$x_0=0.5R_{sh}$ is chosen throughout the paper.
The two earliest time-steps depicted fall in the ejecta-dominated stage, while the others describe the early, intermediate and late Sedov--Taylor stages. 
The top and bottom panels correspond to two injection efficiencies: $\xij=$3 and 4, i.e., $\eta\simeq 2.5\times10^{-3}$ and $\eta=4.1\times10^{-6}$, respectively.

There are many points worth noticing in discussing figure \ref{fig:multi}.
\begin{itemize}
\item As expected, the energy channeled into CRs scales with the number of injected particles: the smaller $\xij$ (larger $\eta$), the larger the energy in non-thermal particles, as can be inferred by comparing the normalization of the curves in the two panels, at any time.
\item As a consequence of the larger CR pressure, the case $\xij=3$ provides a significantly larger magnetic field amplification, and therefore the achievement of a significantly larger maximum energy in the early evolutive stages.
More precisely, in the efficient case (top panel) the spectra turn out to be cut-off around $3-5\times 10^{6}$ GeV, in nice agreement with the position of the knee of the diffuse spectrum of Galactic CRs. 
\item For a given injection efficiency, the CR spectra are almost constant during the ejecta-dominated stage and extend up to the maximum energy achievable during the SNR life.
Nevertheless, when the shock begins to slow down appreciably because of the inertia of the swept-up mass, the instantaneous maximum energy starts to decrease.
The other main effect of the velocity drop is the reduction of the Alfv\'enic Mach number, which leads to steeper and steeper spectra.
The CR acceleration efficiency, however, remains almost constant as long as the shock is still strong ($M_{s,0}\gtrsim10$ up to about $10^4$yr, as depicted in figure \ref{fig:Hydro}).
Such a trend at intermediate ages is non-trivial, and is a consequence of the dependence on the fluid velocity of the streaming instability saturation (see eq.~\ref{eq:PB}).
In other words, the effective Alfv\'en velocity induced by magnetic field amplification decreases more slowly than the shock velocity, therefore leading to a smaller Alfv\'enic Mach number and, in turn, to smaller compression ratios felt by the accelerated particles.
The steepening of the spectra and the decrease of $E_{max}$ turn out to be actually compensated by an increase of the number of particles at lower energies, say around 1--10 GeV.
At the end of the Sedov--Taylor stage also the sonic Mach number drops, $P_{cr}$ decreases almost linearly with time and the spectra become even steeper.
\item Despite of what predicted by standard NLDSA theories, the more efficient the CR acceleration, the steeper the spectra of the accelerated particles (compare the cases $\xij=3$ and 4 in figure \ref{fig:multi}). 
This effect is a consequence of our assumption of calculating the velocity of the scattering centers in the amplified magnetic field rather than in the background one, and can be understood in terms of the effectiveness of the magnetic field amplification.
We will comment more widely on this point in section \ref{sec:injection}.
\end{itemize}

\begin{figure}
\begin{center}
\includegraphics[width=0.9\textwidth]{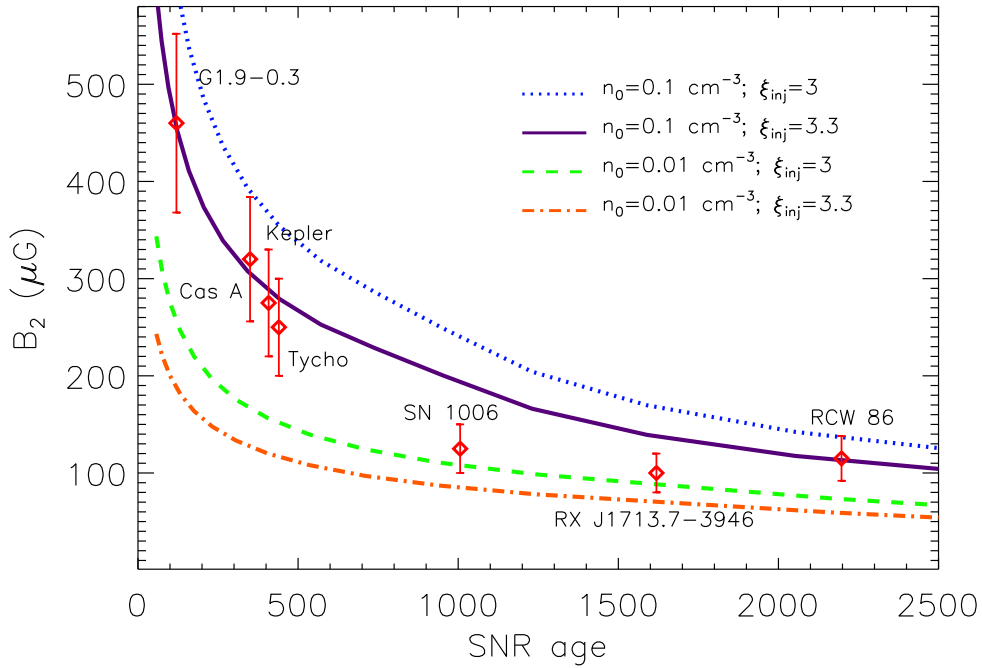}
\caption{Different curves represent the downstream magnetic field produced by CR-induced streaming instability in young SNRs for different injection efficiencies and circumstellar particle densities, as in the legend. 
The data point for historical young SNRs are worked out from the best fitting of X-ray bright rims as due to synchrotron losses, with a fiducial error of 20\% \citep[see refs.][]{V+05,P+06,jumpl,KVB10, Tycho}.}
\label{fig:Bfield}
\end{center}
\end{figure}	

The present findings therefore rely on the magnetic fields to be produced by CR-induced plasma instabilities and more precisely on the presence of modes propagating upstream with respect to the fluid with an effective phase velocity close to the Alfv\'en speed in such enhanced fields.

It is important to stress that, for the more efficient case $\xij=3$, the self-generated magnetic fields that produce such a steepening are consistent with the ones inferred by X-ray observations.
In figure \ref{fig:Bfield} the predicted magnetic fields for different injection efficiencies and circumstellar densities are in fact compared with the downstream fields in very young Galactic SNRs, as inferred  by interpreting the width of their X-ray-bright rims as due to synchrotron losses \citep{V+05,P+06,jumpl,KVB10, Tycho}.

It is worth recalling that the NLDSA theory is actually able to predict the CR acceleration efficiency and the pressure in magnetic fields \emph{as fractions of the bulk pressure $\rho_0V_{sh}^2$}, therefore the absolute value of $B_2$ is rather dependent on our knowledge of the SNR hydrodynamics and of the circumstellar density.
This effect is shown in figure \ref{fig:Bfield}, where a larger $n_0$ naturally leads to a larger $B_2$, at fixed $\xij$, while at fixed particle density the magnetic field increases with the injection efficiency. 
This degeneracy can be broken only if detailed information about $n_0$ and shock velocity is available through a model of the SNR evolution, which in turn requires knowing its age and distance, and/or through the detection of some hadronic emission, which is related to the absolute value of $\rho_0$.
SNRs for which all of these constraints are available are very rare: at the moment Tycho's SNR represents the best case in this respect \citep{Tycho}. 

\begin{figure}
\begin{center}
\includegraphics[width=0.9\textwidth]{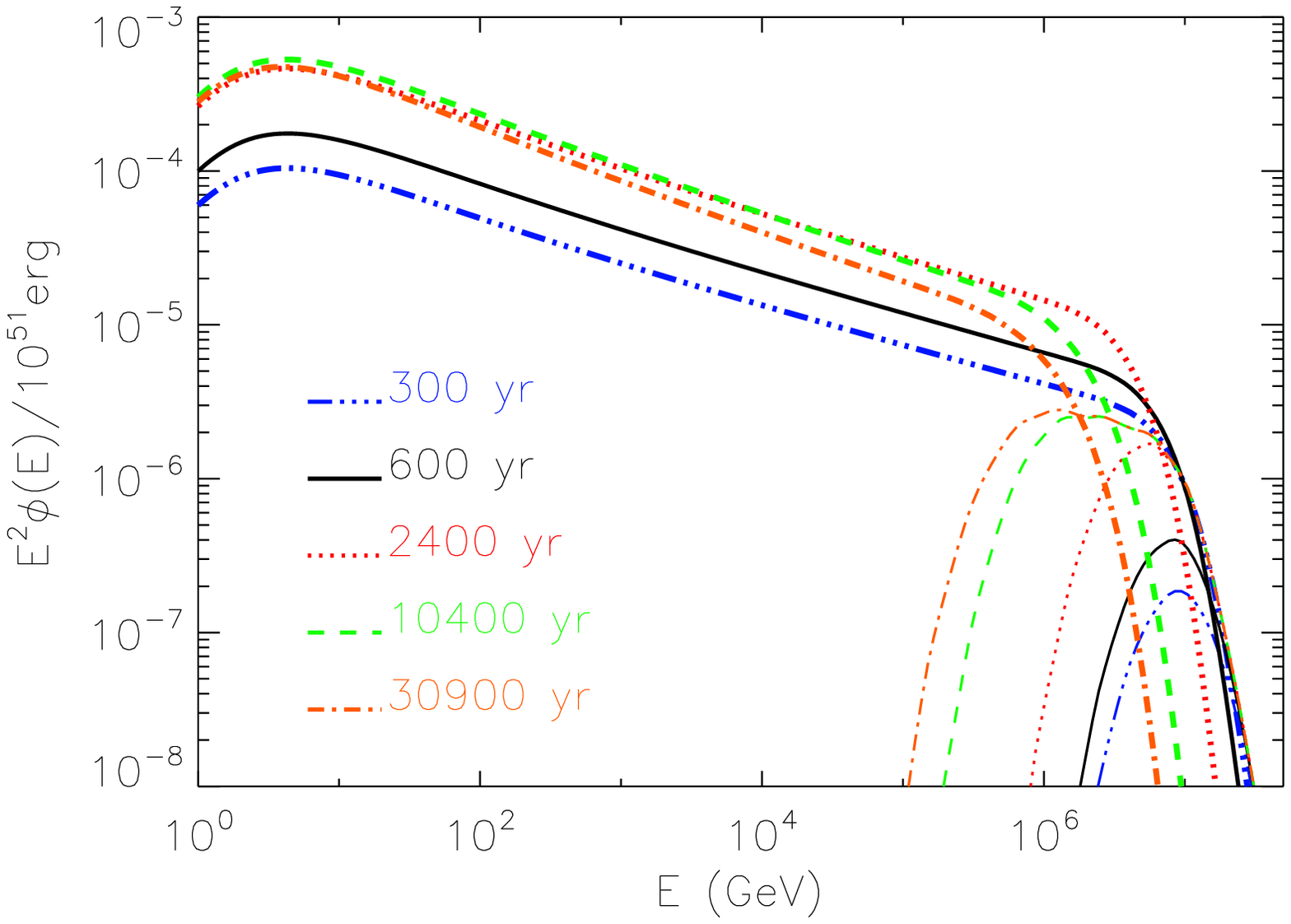}
\includegraphics[width=0.9\textwidth]{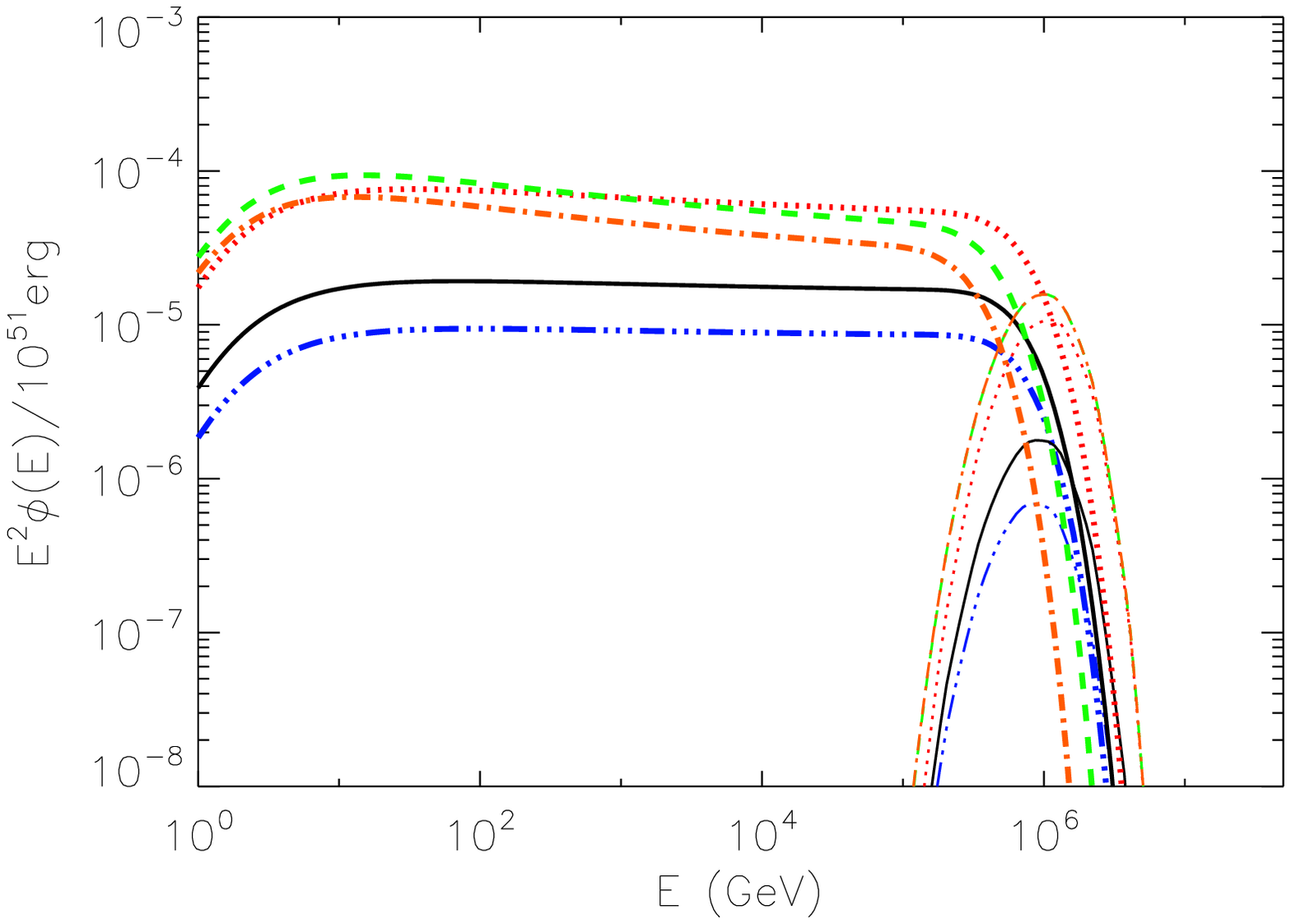}
\caption{Cumulative (i.e., total) CR spectra at different SNR ages as in the legend, divided in advected (thick) and escaped particles (thin lines), for $\xij=3$ and 4 as in figure \ref{fig:multi}.}
\label{fig:multiTOT}
\end{center}
\end{figure}

Some of the effects outlined above are also visible when the cumulative (as opposed to the instantaneous) CR distributions are plotted, as in figure \ref{fig:multiTOT}. 
The thick curves correspond to the total number of CRs advected in the downstream up to the age in the legend, also accounting for the adiabatic losses due to the expansion of the shell (eq.~\ref{eq:adlosses}). 

The thin, bell-shaped curves in figure \ref{fig:multiTOT} show instead the cumulative spectra of the particles which have escaped the SNR from the upstream boundary. 
It is easy to see that escape always occurs at the highest energies, and that the total escape flux grows rapidly during the early stages, reaching a saturation around $\sim 10$ kyr for the parameters chosen here. 
The escape flux is expected to be larger when the CR spectrum is flatter \citep{escape}, and in fact such a trend is consistently recovered in figure \ref{fig:multiTOT}, where the normalization of the escape flux is larger in the bottom panel (case $\xij=4$), despite of the smaller CR acceleration efficiency.

In both cases, the total SNR content in non-thermal particles reaches a sort of saturation after a few thousands years: the contribution of particles accelerated at later and later stages (with steeper spectra) becomes less and less important and is more and more relegated to the lower energies.
At large energies, the contribution is not enough even to balance the adiabatic losses due to the shell expansion: the net effect is a mild steepening of the overall spectrum, visible for both the efficiencies in figure \ref{fig:multiTOT}.
Eventually, adiabatic losses become faster than the particle supply at any energy and the total spectrum normalization start to decrease (compare the $\sim10$ and $\sim 30$ kyr curves).  

Quite interestingly, since the CR acceleration efficiency turns out to be almost constant during the early stages up to about $10^4$yr, some of the predictions put forward in ref.~\cite{crspectrum} are recovered also in the present framework.
More precisely, the convolution over time of the escaped particles turns out to be a power-law spanning the range of variation of $E_{max}$; it especially accounts for the CRs the SNR can inject in the Galaxy above $\sim 10^5$GeV. 
Particles with the same energy which have been advected downstream suffer relevant adiabatic losses, instead: without escape from upstream, SNRs cannot account for the knee observed in the diffuse spectrum of Galactic CRs.

It is very important to stress that, even if the instantaneous spectrum may become steeper than $E^{-2}$ during the Sedov stage, the total CR content of the SNR, which is built up mainly by particles accelerated when the shock is faster, is not expected to immediately reflect such a change. 
In other words, for middle-aged SNRs, a meaningful interpretation of the $\gamma$-ray data requires a time-dependent treatment, in that simple estimates of the SNR content in CRs according to the instantaneous shock velocity and acceleration efficiency may easily lead to both an underestimate of the emission and an overestimate of the spectral slope. 

For these reasons, observing spectra significantly steeper than $E^{-2}$ in the early Sedov stage implies that consistently steep spectra must have been produced since the ejecta-dominated stage: this fact puts an important constraint on the details of how NLDSA works in fast shocks. 
Also for the cumulative CR spectra the same consideration made for the instantaneous spectra holds: the more efficient the CR acceleration, the steeper the resulting spectra and the larger the maximum energy achievable by accelerated particles (compare the case $\xij=3$ and $\xij=4$ in the panels of figure \ref{fig:multiTOT}). 

\subsection{Injection efficiency}\label{sec:injection}
We want to investigate now in more detail the role of the injection parameter in the determination of the expected CR acceleration efficiency.
In particular, it is interesting to check whether the steepening of the spectra which has been observed for low $\xij$ (large $\eta$) may lead to a saturation of the pressure in CRs at the shock.
The opposite limit, the one with large $\xij$ (small $\eta$) is somehow less interesting, since it should simply recover the test particle prediction and therefore lead to very inefficient magnetic field amplification.

Let us consider a fixed SNR age of 3500 yr (early Sedov--Taylor stage), and let $\xij$ vary between 2.5 and 4.5, corresponding to the injection of a fraction between $5\times 10^{-7}$ and $5\times 10^{-2}$ of the particles crossing the shock at any time.  

\begin{figure}
\begin{center}
\includegraphics[width=0.9\textwidth]{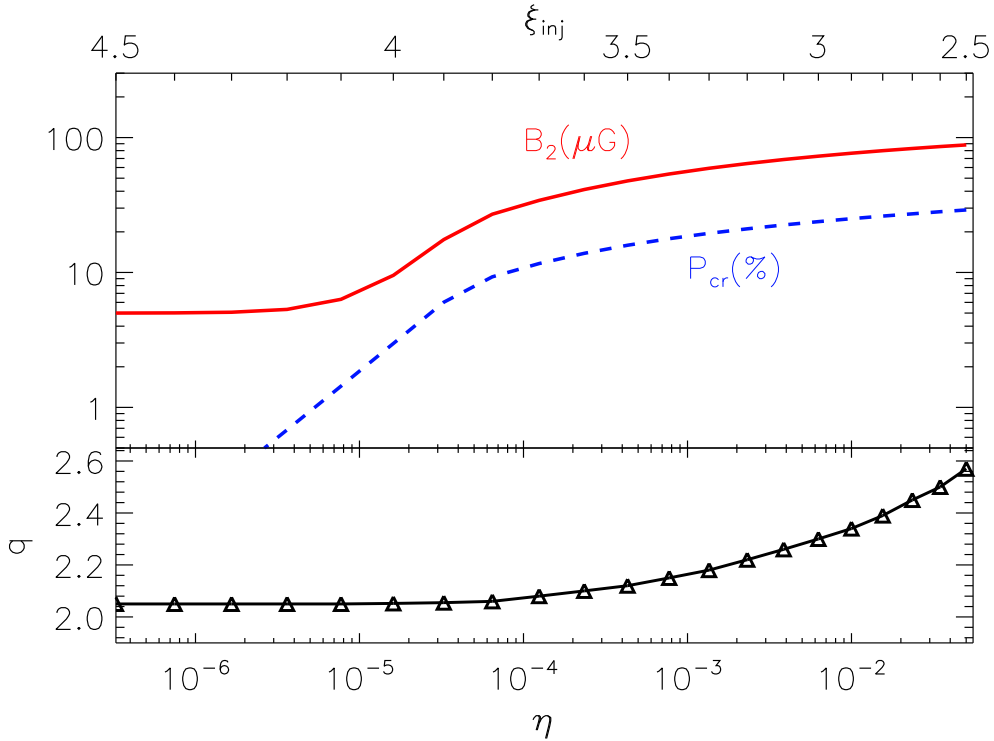}
\caption{Top panel: downstream magnetic field ($B_2$, solid red line) and CR pressure at the shock ($P_{cr}$, dashed blue line) as a function of the fraction of injected particles $\eta$ (see also the corresponding values of $\xij$, green triangles). 
Bottom panel: spectral slope of the particle distribution $q$, always as a function of $\eta$. 
All the quantities are calculated at the same SNR age, 3500 yr, when there is no much difference between the slopes of the instantaneous and cumulative CR distributions.}
\label{fig:eff-VAN+BC}
\end{center}
\end{figure}

In the top panel of figure \ref{fig:eff-VAN+BC} the fraction of the bulk pressure in CRs, $P_{cr}=P_{cr,2}$, and the downstream magnetic field, $B_2$, are shown as a function of the fraction of injected particles. 
The value of $\xij$ that needs to be chosen to inject the correspondent fraction $\eta$ is showed in the top axis.  

For $\eta$ smaller than $\sim 10^{-5}$ the acceleration efficiency is less than a few \%, and in turn magnetic field amplification is very inefficient: downstream we are basically left with the component parallel to the shock normal only: $B_0\approx 5\mu$G.
From $\eta\simeq 10^{-6}$ to $\eta\simeq 10^{-4}$ $P_{cr}$ increases almost linearly (up to almost 10\%) and so does the magnetic field.

The most interesting effect, though, is that for $\eta$ larger than $10^{-4}$ the pressure in CRs remains almost constant, between 10 and 30\%.
The variation of almost three orders of magnitude in $\eta$, in fact, reflects in a change of less than a factor three in $P_{cr}$ and $B_2$.
The very reason for this effect is accounted for in the bottom panel of figure \ref{fig:eff-VAN+BC}, where the spectral slope of the CR distribution at the shock 
$q=-\frac{d\log{\phi(E)}}{d\log{E}}$ is shown.
When the shock modification reaches about 10\%, the feedback of the finite velocity of the scattering centers kicks in, leading to a steepening of the spectrum with respect to the $\propto E^{-2}$ prediction of the test-particle limit.
The less and less energy that goes into the high-energy part of the CR spectrum (which extends to higher energies, though) is rather fairly balanced by the more and more energy channeled into the lower energy particles, in such a fashion that the overall pressure and density energy in CRs turn out to be rather insensitive to the injection details.
Nevertheless, the spectral slope of the CR distribution increases with the increase of the injection efficiency up to $q\simeq 2.6$, spanning values consistent with most of the $\gamma$-ray bright SNRs \citep{gamma}.

Such an effect gives a new flavor to the notion of a \emph{CR-modified shock}: an \emph{efficient CR acceleration} does not lead to an arbitrary increase of the pressure in CRs and therefore to spectra asymptotically as flat as $\propto E^{-1.2}$ \citep[see, e.g.,][and references therein]{ab06}, but rather to a significant steepening of the particle spectra with a moderate fraction (10--30\%) of the bulk energy channeled into non-thermal particles. 
This is probably the most interesting original finding of the present work, and represents an important step in order to consistently account for the new wealth of information coming from $\gamma$-ray observations and the long-lasting SNR paradigm for the origin of Galactic CRs. 

We want also to stress that other mechanisms as turbulent heating \citep[see e.g.,][]{mckenzie-volk82,berezhko-ellison99} and dynamical magnetic feedback \citep{jumpl} have been proposed in order to avoid the arising of very modified shocks, and consequently the appearing of very concave spectra as flat as $E^{-1.2}-E^{-1.5}$ at the highest energies, which would be at odds not only with $\gamma$-ray but also with radio and X-ray observations of SNRs. 
The first mechanism is however unlikely to be very effective, in that it would require a severe dissipation of the magnetic turbulence, contrary to the several evidences of amplified magnetic fields found in young SNRs.
The latter, instead, is expected to be rather relevant in the current framework (and it is in fact included in the present calculation) and may effectively contribute to preventing an excessive shock modification \citep{jumpkin}. 

However, neither the first nor the latter piece of Physics, alone or together, can account for the required steepening of the CR spectra in that, if one does not allow for a distinct phenomenology of the compression ratios felt by the fluid and the CRs, the predicted spectral slope cannot in any case be larger than 2 for strong shocks at energies above a few GeV, i.e., in the region of interest for $\gamma$-ray astronomy.

\subsection{Some additional comments}\label{sec:comments}
A natural question about the findings illustrated in the previous section is how strong is the dependence of the steepening on the (still) not-well understood nature of the magnetic turbulence.
Quantitatively speaking, for the present mechanism to be effective, the magnetic irregularities need to have a phase velocity not-negligible with respect to the fluid one, on a length-scale comparable with the diffusion length of the most energetic particles.
$V_A/U\simeq 10-15\%$, corresponding to effective $M_A\simeq 6-8$, is enough to return a spectral slope of 2.2--2.35.
The sign of the correction in the compression ratios (eq.~\ref{eq:rsrtt}) comes for free when the turbulence is generated by the CR streaming, since upstream waves should propagate against the CR gradient, i.e., against the fluid, while downstream it seems unlikely to have the propagation of waves with any given helicity since the medium is hot and turbulent.

Another fundamental property required for the overall steepening of the CR spectra is that also particles diffusing far in the precursor must scatter against magnetic structures moving with a phase velocity comparable to the ones of modes propagating closer to the subshock. 
It is in fact possible to show that by assuming a local magnetic field amplification as given by eq.~(\ref{eq:PB}) as a function of $x$, which would imply less and less turbulence closer and closer to the free-escape boundary, the CR spectrum remains as flat as $E^{-2}$ close to the maximum energy.
In other words, either energetic particles diffusing up to the upstream edge of the precursor decouple from the fluid and escape the system (in a fashion more complex than the one described by the transport equation \ref{eq:trans}), or there needs to be enough ongoing magnetic field amplification even rather far upstream, on the scales relevant to the highest energy particles. 

Whether such a magnetic configuration is really achieved in SNR shock cannot be addressed with analytical techniques, since it requires the investigation of very complex particle--wave interactions, well beyond the quasi-linear limit.
The most useful insights into these details come instead from PIC simulations, in which the problem is tackled from basic principles \citep[see e.g.,][]{stroman+09,rs09,rs10}.
 
Bell modes are expected to provide an additional channel for amplify the magnetic field, especially close to the escape boundary where the CR distribution function is rather anisotropic. 
As pointed out in refs.~\cite{stroman+09,rs09}, while in the linear regime of the instability modes are almost purely growing, the saturation of Bell modes in the non-linear regime ($\delta B/B_0 \sim 10$) may occur exactly because of the local Alfv\'en velocity to become comparable with the drift velocity of the CR current.

An additional promising class of Alfv\'enic modes is represented by the left-hand, circularly polarized ion-whistler waves found in hybrid (kinetic protons/fluid electrons) simulations described in ref.~\cite{gs12}. 
In their work, in fact, the authors find that, in the upstream parallel shocks with $M_A\gtrsim 10$, modes with phase velocity larger than the Alfv\'en velocity can be efficiently excited.   

Finally, the possibility of producing large-wavelength modes through the fire-hose mechanism has been put forward by some authors when investigating the development of the beam instability \cite{gary+84,shapiro+98}.

Even if a lot of work has still to be done in order to understand which instability is the most relevant for SNR shocks, these scenarios are quite consistent with our working hypothesis that, when some channel for CR-induced magnetic field amplification is active and leads to sizable perturbations with $\delta B/B \gtrsim 1$, the wave--particle coupling tends to reduce the gap between the velocities of the magnetic structures and of the CR drift: the net effect is therefore to force the magnetic structure to acquire a non-negligible velocity with respect to the fluid, as heuristically implemented in section \ref{sec:model}.

From the point of view of the phenomenology of $\gamma$-ray bright SNRs and their steep spectra, it is important to recall that there is another mechanism which may be relevant to reconcile theory and observations.
As pointed out in ref.~\cite{neutri}, when a shock propagates into a partially ionized medium, charge-exchange processes between neutral and protons in the downstream lead to the formation of a \emph{neutral return flux} which eventually conveys some of the downstream energy back in the upstream, producing a neutral-induced precursor on a scale comparable with the ionization/charge-exchange mean free path ($\sim 10^{17}$ cm for typical SNR parameters). 
Such a \emph{neutral-induced precursor}, which may reduce the sonic Mach number down to $2-3$ before the subshock, develops on a scale independent of the CR diffusion length, therefore can in principle induce a sizable modification in the velocity profile and in turn on the expected compression ratios, possibly leading to steep CR spectra up to the highest energies.

The only limitation of this unavoidable mechanism is that it is expected to be relevant only when the shock velocity is less than about 3000 km/s, because when the relative velocity between a neutral and a proton (which immediately downstream is of order of $V_{sh}$) is larger than this critical value ionization becomes favored with respect to charge-exchange.
Therefore, while during the Sedov--Taylor stage the neutral return flux has to be reckoned with both in working out SNR hydrodynamics and CR acceleration, it is unlikely for such a phenomenon to produce steep spectra during the ejecta-dominated stage as well.

A further comment is due about the choice of the environmental parameters adopted in the paper. 
It is rather clear that the Alfv\'enic Mach number $M_{A,0}\propto B_0/\sqrt{\rho_0}$, which plays a crucial role in the matter at debate, may differ quite a bit according to the nature of the circumstellar medium the shock propagates into.
In superbubble-like environments, for instance, the particle density may be as low as $\sim 10^{-3}{\rm cm}^{-3}$, therefore the effects discussed here may be even more marked.
On the other hand, one could argue that for a type-Ia SN the shock should propagate in a denser and colder medium with density of order 0.1--1cm$^{-3}$, in turn implying significantly larger $M_A$.

These arguments, however, must be taken with a grain of salt, since the dependence on the plasma density may easily be overbalanced by a more or less efficient magnetic field amplification, which in many cases is expected to be mildly dependent on $B_0$.
The \emph{absolute} value of the downstream magnetic field, then, may be rather dependent on the actual shock velocity and on the circumstellar density, as discussed in commenting figure \ref{fig:Bfield}, therefore the analysis of a given remnant may differ quite a bit from this general outline.
In the case of Tycho's SNR, for instance, assuming a particle density $\sim 0.3{\rm cm}^{-3}$ and $\xij=3.7$ does not lead to results qualitatively much different from the ones worked out here: also in that case, in fact, the inferred magnetic fields are strong enough to produce an appreciable steepening of the proton spectrum \citep{Tycho}.

If any, the actual value of $M_{A,0}$ may be very important in the regions  where magnetic field amplification may not be maximally efficient (like far upstream), therefore regulating the steepness of the spectrum up to the highest energies.
In this respect, it is easy to show that for hot, rarefied cavities in bubble-like environments ($T_0\sim 10^6$--$10^7$, $n_0\sim 10^{-3}{\rm cm}^{-3}$) with standard Galactic fields $B_0\sim 5\mu$G both the sonic and the Alfv\'enic Mach numbers become of order 5--10 well within the Sedov stage.   

\section{Conclusions}\label{sec:conclusions}
In this paper we tackled the problem of particle acceleration at SNR shocks in order to provide a theoretical explanation for the many evidences of spectra steeper than $E^{-2}$ coming from $\gamma$-ray observations of SNRs and from our current understanding of CR propagation in the Galaxy.

The present investigation is motivated by the fact that the prediction of the most natural NLDSA theory, namely that the larger the CR acceleration efficiency, the flatter the CR spectrum, needs to be revised in order to be consistent with current observations.

We demonstrated that the magnetic field amplification naturally induced by the super-Alfv\'enic streaming of accelerated particles may significantly alter the properties of the CR scattering, actually decoupling the compression ratios felt by the fluid and by the diffusing particles.
Under reasonable assumptions about the development and the saturation of the plasma instabilities (which still need to be checked against first-principle simulations given the impossibility of carrying out an analytical treatment of their non-linear regime), we find that the self-generated magnetic turbulence can lead to a steepening in the spectrum of the accelerated particles.

More precisely, the expected levels of magnetic field amplification granted by streaming instabilities, in addition to be consistent with the fields inferred in the downstream of young SNRs (figure \ref{fig:Bfield}), can lead to CR spectra as steep as $\sim E^{-2.3}$ even in the early stages of the SNR evolution, i.e., when the sonic Mach number is still much larger than one.

The effective steepening of the CR spectra turns out to be function of the CR acceleration efficiency, which we tune by regulating the fraction of particles extracted from the thermal bath and injected in the acceleration process, $\eta$. 
Such a dependence is however radically different from the one predicted by a NLDSA theory in which the finite velocity of the scattering centers is not taken into account: the spectra of the accelerated particles, in fact, is showed to be consistent with the test-particle prediction for very low efficiencies, but invariably steeper and steeper than $E^{-2}$ when $\eta$ becomes larger and larger.

Very interestingly, in this non-linear system a larger $\eta$ on one hand produces a larger $P_{cr}$, but on the other hand it also produces a larger self-generated magnetic field, which acts in such a way to reduce the shock modification by steepening the CR spectrum.
The net effect, depicted in figure \ref{fig:eff-VAN+BC}, is that for $\eta\gtrsim 10^{-4}$
the pressure in CRs saturates around 10--30\% of the shock bulk pressure; 
also the value of the self-generated magnetic field saturates in a similar fashion (figure \ref{fig:Bfield}). 
This self-regulating interplay between efficient CR acceleration and effective magnetic field amplification may represent a key ingredient in order to quantitatively explain both the levels of magnetization inferred in SNRs and the CR acceleration efficiency required for SNRs to be the sources of Galactic CRs.

Another important result is that the magnetic fields inferred are also the ones required to accelerate particles up to energies comparable with the observed knee in the diffuse spectrum of Galactic CRs, namely about 3--5$\times 10^{6}$GeV for protons (top panel of figure \ref{fig:multiTOT}). 

We also investigated the temporal evolution of the CR spectra following the SNR during its ejecta-dominated and adiabatic stages, assessing the differences between instantaneous and cumulative spectra of the particles advected downstream (which undergo adiabatic losses due to the shell expansion), and of the particles escaping the system from the upstream as a consequence of the decrease of the SNR confining power with time \citep{escape}.
We find that the CR acceleration efficiency is expected to drop with time, basically because of the slowing down of the shock due to the inertia of the swept-up material, with the net result that in the intermediate/late Sedov phase the SNR content in CRs is invariably dominated by the contributions of the earlier stages.

This is particularly important for two reasons.
First, it shows that, in order for the total CR contribution during the SNR lifetime to be steeper than $E^{-2}$, particles must be accelerated with steep spectra also during the early stages. 
Second, when calculating the expected non-thermal emission from middle-age SNRs, a time-dependent study of the SNR evolution has to be carried out, since a simple snapshot of what is going on at the shock would easily lead to an overestimate of the spectral slope and to a underestimate of the SNR content in non-thermal particles.

\acknowledgments
I would like to thank A. Spitkovsky and P. Blasi for their constant advices and support, D. Giannios, B. Metzger, E. Amato, G. Morlino and S. Funk for having read a preliminary version of this paper and an anonymous referee for his/her comments. 
This research work was supported by NSF grant AST-0807381.

\bibliographystyle{JHEP}
\bibliography{bibeff}

\end{document}